\documentclass[rsc,twocolumn,showpacs]{revtex4-1}
\usepackage[dvips]{graphicx}
\usepackage{amssymb}
\usepackage{amsmath,amsthm,amsfonts,amssymb,bm}
\usepackage{multirow}
\usepackage{hyperref}

\begin{document}

\title{Long-lived N\'{e}el states in antiferromagnetic quantum spin chains with strong uniaxial anisotropy for atomic-scale antiferromagnetic spintronics}
\author{Jun Li}
\author{Bang-Gui Liu}
\email[Corresponding author:~]{bgliu@iphy.ac.cn}
\affiliation{Beijing National Laboratory for Condensed Matter
Physics, Institute of Physics, Chinese Academy of Sciences, Beijing
100190, China}

\begin{abstract}
It has been experimentally established that magnetic adatoms on surfaces can be arranged to form antiferromagnetic quantum spin chains with strong uniaxial anisotropy and Neel states in such spin systems can be used to realize information storage. Here, we investigate eigen states, quantum spin dynamics, and life times of Neel states in short antiferromagnetic quantum spin chains with strong uniaxial anisotropy on the basis of numerical exact diagonalization method. We show rigorously that as long as the uniaxial anisotropy is very strong, the ground state and the first excitation state, being nearly degenerate, are safely separated from the other states and thus dominate the quantum dynamics of the Neel states. Through further numerical analysis, we achieve a powerful life-time expression of the Neel states for arbitrary spin and model parameters. It is interesting that for the famous Fe adatom chains on Cu$_2$N surface, 14 or 16 Fe adatoms are enough to obtain a practical long life-time for Neel state storage of information. These should be applicable to other similar antiferromagnetic spin systems for atomic-scale antiferromagnetic spintronics.
\end{abstract}


\maketitle

\vspace{0.5cm}
{\noindent \Large 1. Introduction}
\vspace{0.5cm}


It is inspiring that adatom-based antiferromagnets have been realized on semiconductor surfaces and used for novel magnetic information storage because their N\'{e}el states can be stabilized by strong uniaxial single-ion magnetic anisotropy\cite{ATM1,ATM2,ATM3}. Such nanomagnets can be fabricated adatom by adatom, and their spin anisotropy can be controlled\cite{AFM3,AFM4,AFM1,AFM2}. Thus, one can make  antiferromagnetic chains, bi-chains, nano-ribons, or nano-sheets consisting of several or tens of adatom spins with strong magnetic anisotropy and adjustable inter-spin interactions. Spin chains are of much interest because they belong to an important category of Heisenberg spin models. In fact, various one-dimensional antiferromagnetic Heisenberg models have been intensively investigated\cite{AHM2,B-S,AHM4,AHM6,AHM7,TIME1,THEORY5,AHM5}. For $S=1$, there exists an inetresting Haldane topological phase if there is no strong uniaxial magnetic anisotropy\cite{AHM3,Haldane,AHM9}. On experimental side, one usually use high spins with strong uniaxial single-ion magnetic anisotropy in adadtom spin systems\cite{ATM1,ATM2,ATM3,AFM3,AFM4,AFM1,AFM2}. It is known that strong uniaxial single-ion anisotropy is necessary to achieve stable N\'{e}el states. Experimentally, electrons currents injected through STM tips have been used to  control the N\'{e}el states for information storage\cite{ATM3}.
On theoretical side, some efforts have been made to understand and explore controlling the adatom-spin antiferromagnets with spin-polarized electron current\cite{THEORY1,lj0}, spin current\cite{THEORY4}, and mechanical oscillator\cite{THEORY2} and to investigate symmetry effects on spin switching of single adatoms\cite{THEORY3}. It is believed that more significant advances and deeper insight in this field can likely lead to an atomic-scale antiferromagnetic spintronics.

Here, we investigate the intrinsic quantum dynamics and life times of N\'{e}el states in the quantum Heisenberg antiferromagnetic chain model consisting of $2N$ spins ($S\ge 1$) with strong uniaixal single-ion anisotropy. We accurately calculate eigenvalues and eigenfunctions through exact diagonalization, and rigorously show that the ground state and the first excitation can be both safely separated from the other states and well described with the two N\'{e}el states as long as the single-ion anisotropy is very strong. Then, we thereby calculate the switching rates and life times of the N\'{e}el states. Surprisingly, we achieve a unified powerful expression of the life times through fitting our accurate numerical results. More importantly, for the Fe-adatom spin antiferromagnets on Cu$_2$N semiconductor surface\cite{ATM1,ATM2,ATM3}, $2N$= 14 or 16 is large enough to achieve practical life times of N\'{e}el states for information storage. More detailed results will be presented in the following.

\begin{figure*}[!htbp]
  \includegraphics[width=14cm]{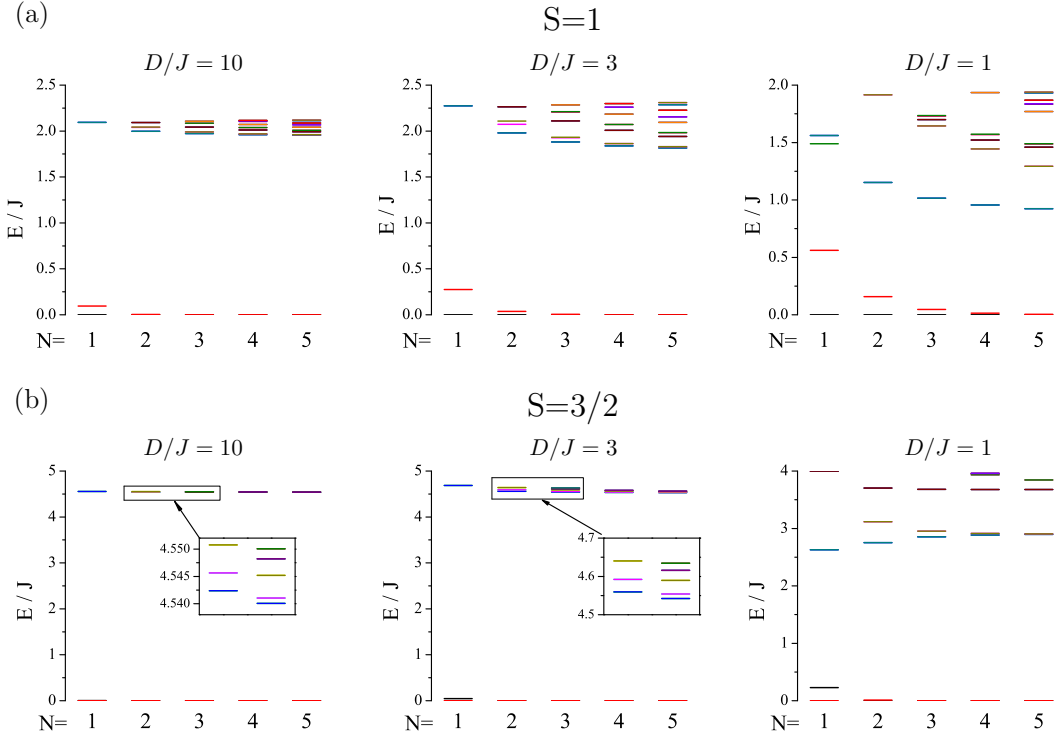}
 \caption{The eigen energies ($E/J$) of Hamiltonian (\ref{eq:1}) for $S=1$ (a) and $S=3/2$ (b), with $D/J=10$, 3, and 1.  } \label{fig1}
\end{figure*}

\begin{table*}[!htbp]
 \centering
 \caption{ The $N$-dependent energy gap $\Delta E$ between the ground state G and the first excitation E$_1$ for two spin values (1 and 3/2) and three $D/J$ (1, 3, and 10). }\label{tab1}
 \begin{tabular}{c|c|c|c|c|c|c}
   \hline
         &       &$N=1$       & $N=2$        & $N=3$     & $N=4$    & $N=5$\\
   \hline
$S=1$&$D/J=1$    &$0.561$      &$0.158 $        &$4.76 \times10^{-2}$ &$1.43 \times10^{-2}$     &$4.32\times10^{-3}$        \\
   &$D/J=3$      &$0.275$      &$3.66\times10^{-2}$        &$4.95\times10^{-3}$     &$6.68\times10^{-4}$    &$9.03\times10^{-5}$     \\
   &$D/J=10$     &$9.48\times10^{-2}$         &$4.41\times10^{-3}$        &$2.05\times10^{-4}$      &$9.56\times10^{-6}$     &$4.46\times10^{-7}$        \\
   \hline
$S=3/2$&$D/J=1$&$0.228$ &$1.24\times10^{-2}$ &$6.64\times10^{-4}$&$3.56\times10^{-5}$&$1.90\times10^{-6}$        \\
   &$D/J=3$      &$4.53\times10^{-2}$      &$4.44\times10^{-4}$        &$4.34\times10^{-6}$     &$4.25\times10^{-8}$    &$4.15\times10^{-10}$     \\
   &$D/J=10$     &$5.09\times10^{-3}$      &$5.69\times10^{-6}$        &$6.36\times10^{-9}$     &$7.96\times10^{-12}$   &$8.81\times10^{-15}$        \\
   \hline
 \end{tabular}
\end{table*}

\vspace{0.5cm}
{\noindent \Large 2. Results and discussion}
\vspace{0.5cm}

\vspace{0.25cm}
{\noindent \large 2.1 Spin Model and eigenstates}
\vspace{0.25cm}

We start with general
one-dimensional quantum Heisenberg antiferromagnetic model with strong
uniaxial single-ion anisotropy,
\begin{equation}
\hat{H}=J\sum_{i=1}^{2N-1}\hat{\vec{S}}_{i}\cdot \hat{\vec{S}}_{i+1}
-D\sum_{i=1}^{2N}(\hat{S}_{i}^{z})^{2}, \label{eq:1}
\end{equation}
where the total number of the spins is $2N$, the parameter $J$ ($>0$) is the
antiferromagnetic exchange constant, $D$ ($>0$) is
used to characterize the single-ion magnetic anisotropy in the z
axis, and $\hat{\vec{S}}_{i}$ is the spin operator at site $i$, satisfying open boundary condition. Here, we do not need any inhomogeneous effective magnetic field to split the two N\'{e}el states\cite{ATM3,THEORY1}, but to experimentally prepare a specific N\'{e}el state, one can use an STM tip to inject a spin-polarized electron current on the first adatom spin\cite{ATM3}. If being applied to similar spin rings with uniaxial anisotropy, such as antiferromagnetic molecule wheels\cite{spinwheel}, the Hamiltonian (1) needs some modification to make the spin operators satisfy periodic boundary condition.

Using $\hat{S}^{\pm}_i=\hat{S}^{x}_i\pm i\hat{S}^{y}_i$, we have
$\hat{\vec{S}}_{i}\cdot \hat{\vec{S}}_{i+1}=\hat{S}_{i}^{z}\cdot
\hat{S}_{i+1}^{z} + \frac{1}{2} (\hat{S}_{i}^{+}\cdot
\hat{S}_{i+1}^{-}+\hat{S}_{i}^{-}\cdot \hat{S}_{i+1}^{+})$. The ideal N\'{e}el states ($|N_1\rangle$ and $|N_2\rangle$) are certainly not the eigenstates of Hamiltonian (\ref{eq:1}) due to the transverse part including the
raising and lowing operators $\hat{S}^{\pm}_i$, but the strong
single-ion anisotropy $D$ in the z axis makes the spin tend to orient in
the z axis. Consequently, there are large ($D/J$ dependent) weight of the
two N\'{e}el states in the ground state and low excitations,
which implies that the two N\'{e}el states for large $D/J$ can be stable enough
to be used for information storage\cite{ATM3}. Using exact
diagonalization method\cite{spin1} to the Hamiltonian (1), we can obtain the spin
eigenvalues and eigenfunctions. For convenience, we shall
use $J$ as our unit in the following, which means that anisotropy parameter $D$ and
energy $E$ can be scaled in terms of $J$.

In Fig. 1 we present the energy eigenvalues depending on $N$ (1
through 5) and $D/J$ (10, 3, and 1) for $S=1$ and $3/2$. For each of
the cases, the ground state G and the first excitation E$_1$ are both
separated from the other states. The trend is that the separation
increases with $D/J$ and $S$. The corresponding energy gaps between the ground
states and the first excitation ones are summarized in Table
\ref{tab1}. It is clear in the table that the gap decreases with
$N$, $S$, and $D/J$. The weights of the N\'{e}el states in the ground
state (G) and the first excitation (E$_1$) as functions of $N$ are
presented in Table \ref{tab2} for the two spin values and the three $D/J$
ones. It can be seen that the N\'{e}el weights increase with $D/J$,
but decrease with $N$. Except the special case of $N=1$, the
N\'{e}el weights increase with $S$, too. Because we are interested in the cases with strong uniaxial anisotropy, the ground state and low excitations are far from the regime of the Haldane state\cite{AHM3,Haldane,AHM9}.

\begin{table}[!h]
 \centering \caption{The $N$-dependent N\'{e}el weights in the ground state (G) and the first excitation (E$_1$) for different $S$ and $D/J$. }\label{tab2}
 \begin{tabular}{c|c|c|ccccc}
   \hline
       &       &    &$N=1$     & $N=2$    & $N=3$    & $N=4$    & $N=5$\\
   \hline
$S=1$  &$D/J=1$  & G  &$0.864$   &$0.750$   &$0.693$   &$0.634$   &$0.575$     \\
       &       & E$_1$ &$1.000$   &$0.854$   &$0.735$   &$0.650$   &$0.580$     \\
       &$D/J=3$  & G  &$0.964$   &$0.935$   &$0.916$   &$0.892$   &$0.867$     \\
       &       & E$_1$ &$1.000$   &$0.958$   &$0.921$   &$0.893$   &$0.868$     \\
       &$D/J=10$ & G  &$0.996$   &$0.991$   &$0.988$   &$0.985$   &$0.980$     \\
       &       & E$_1$ &$1.000$   &$0.994$   &$0.989$   &$0.985$   &$0.980$     \\
   \hline
$S=3/2$&$D/J=1$  & G  &$0.900$   &$0.875$   &$0.822$   &$0.770$   &$0.720$     \\
       &       & E$_1$ &$0.968$   &$0.882$   &$0.823$   &$0.770$   &$0.720$     \\
       &$D/J=3$  & G  &$0.985$   &$0.946$   &$0.957$   &$0.941$   &$0.926$     \\
       &       & E$_1$ &$0.991$   &$0.973$   &$0.957$   &$0.941$   &$0.926$     \\
       &$D/J=10$ & G  &$0.999$   &$0.995$   &$0.994$   &$0.992$   &$0.991$     \\
       &       & E$_1$ &$0.999$   &$0.996$   &$0.994$   &$0.992$   &$0.991$     \\
   \hline
 \end{tabular}
\end{table}

It is easy to prove that the total spin z-component
$\hat{S}^z=\sum_i\hat{S}^z_i$ is conserved because it is commutable with
the Hamiltonian (1). All the energy eigenstates can be classified in
terms of the eigenvalue $S^z$ of $\hat{S}^z$. Generally speaking,
for a finite antiferromagnetic chain with $2N$ spins, the ground
state is a spin single state. When $D/J$ is very large, the ground
state can be approximately constructed with a superposition of the
two N\'{e}el states. For general $D/J$, we can always construct the
following two eigenstates from the N\'{e}el states.
\begin{equation}\left\{
\begin{array}{c}
|+\rangle=c_{1}(|N_{1}\rangle+|N_{2}\rangle
+O_{1}|\mathcal{O}_{+}\rangle+\cdots)\\
|-\rangle=c_{2}(|N_{1}\rangle-|N_{2}\rangle
+O_{2}|\mathcal{O}_{-}\rangle+\cdots),
\end{array}\right. \label{eq:2}
\end{equation}
where $|\mathcal{O}_{\pm}\rangle$ is defined as
$\sum_{i=1}^{2N-1}(\hat{S}^{+}_{i}\cdot
\hat{S}^{-}_{i+1}+\hat{S}^{-}_{i}\cdot
\hat{S}^{+}_{i+1})(|N_{1}\rangle\pm|N_{2}\rangle)$, and $c_1$,
$c_2$, $O_1$, and $O_2$ are coefficients to be determined. Actually,
our exact diagonalization results show that when $2NS$ is even, the
ground state G is $|+\rangle$ and the first excitation E$_1$ is
$|-\rangle$; and when $2NS$ is odd, we have G=$|-\rangle$ and
E$_1$=$|+\rangle$. This is in accordance with the theoretical
results obtained by spin coherent state path integral\cite{FE2,FEN}. The
higher excitation states with E$_i$ ($i\ge 2$) can be constructed in
the similar way.

\vspace{0.25cm}
{\noindent \large 2.2 Quantum dynamics of N\'{e}el states}
\vspace{0.25cm}

 We shall mainly focus on
the subspace of the states with $S^z=0$ because the ground state and
the low excitation states including the N\'{e}el states belong to
this subspace, and however, we shall turn to other states when we
discuss the effect of temperature. For convenience, we shall use
$|g\rangle$ and $|e_i\rangle$ ($i\ge 1$) to denote all the eigenstates in the
$S^z=0$ subspace. Because this subspace is closed under the Hamiltonian (1), the time evolution of the two N\'{e}el states can be expanded as
\begin{equation}
|\tilde{N}_a(t)\rangle=f^a_0e^{iE_0t/\hbar}|g\rangle+\sum_{j\ge
1}f^a_je^{iE_jt/\hbar}|e_j\rangle,
\end{equation}
where $E_j$ and $f^a_j$ ($j\ge 0$, $a=$1,2) are the eigenvalues and
expansion coefficients of the $j$-th eigenstates. Here, of course, we have
$|\tilde{N}_a(0)\rangle=|N_a\rangle$, $|g\rangle$=G, and $|e_1\rangle$=E$_1$. Then, the weight of
$|N_a\rangle$ in $|\tilde{N}_a(t)\rangle$ can be expressed as
\begin{equation}
\chi_a^2(t)=|f^a_0+\sum_{j\ge 1}f^a_je^{i\Delta E_jt/\hbar}|^2,
\end{equation}
where $\Delta E_j=E_j-E_0$. The total N\'{e}el weight of
$|N_1\rangle$ and $|N_2\rangle$ in $|\tilde{N}_1(t)\rangle$ can be
defined as $W_N(t)=\chi_1^2(t)+\chi_2^2(t)$. $W_N(t)$ reflects how well the
N\'{e}el states describe the quantum antiferromagnetic chain.
The two-state approximation results in a simplified expansion of
$|\tilde{N}_a(t)\rangle$, such as
\begin{equation}
|\tilde{N}_1(t)\rangle\propto\cos(\frac{\Delta
E_1}{2\hbar}t)|N_1\rangle+\sin(\frac{\Delta E_1}{2\hbar}t)|N_2\rangle.
\end{equation}
We present $\chi_1^2(t)$ and $W_N(t)$ in Fig. 2 for $D/J$=10 , 3,
and 1. For $\chi_1^2(t)$, the two-state approximation is also
presented for comparison. It is clear that $\chi_1^2(t)$ is a
periodic function of $t$ and $W_N(t)$ is almost a constant except
a narrowly oscillating noise due to the higher states. For small $D/J$
such as 1, the maximal value of $\chi_1^2(t)$ is approximately 0.8 and the
N\'{e}el weight $W_N(t)$ is less than 0.9, but for large $D/J$ such as
10, $\chi_1^2(t)$ can be well described with $\cos^2(t/2T)$ and the N\'{e}el
weight becomes larger than 0.99. Here, the time period is equivalent to $P=2\pi T$, and $1/T$ reflects the switching rate (or frequency) between the two N\'{e}el states. It is surprising that for this case of $S=2$ and
$N=2$, $T$ increases by five orders of magnitudes when
$D/J$ changes from 1 to 10.

\begin{figure}[!htbp]
  \includegraphics[width=8.5cm]{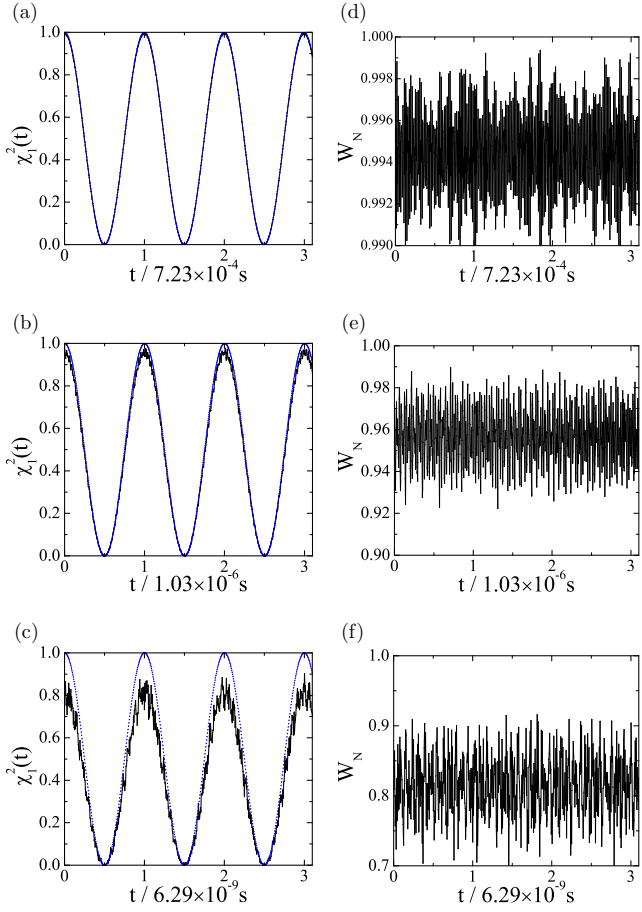}
 \caption{The time dependence of $\chi_1^2(t)$ (black lines in the left column) and $W_N(t)$
(the right column) for $S=2$, $2N=4$, and $D/J=10$ (a,d), 3 (b,e),
and 1 (c,f), respectively. The two-state approximated results of
$\chi_1^2(t)$ (blue dash lines in the left column ) are also
presented for comparison.} \label{fig2}
\end{figure}

\vspace{0.25cm}
{\noindent \large 2.3 Life times of N\'{e}el states}
\vspace{0.25cm}

Because $\chi_1^2(t)$ is a
well-defined periodic function of $t$, the quantity $T$, the time spent by a switching circle between the two N\'{e}el states, can be used to
characterize the life times of the N\'{e}el states. In the case of
two-spin chains ($N=1$) with $S\le 3$, we can calculate eigenstates
and $|\tilde{N}_a(t)\rangle$ exactly. For $S=1$, we obtain $\Delta E_1
= J[\sqrt{4(D/J)^{2}+4D/J+9}-2D-1]/2$, and $T$ can be
expressed as $(2D/J+1)/2$ when $D/J$ is large. For higher
spins, we can achieve $T \propto(2D/J+1)^{2S-1}$ for both integer and
half odd integer spins by using a usual perturbation method.
Generally speaking, we can also use exact
diagonalization method to calculate $\Delta E$ and $T$ for arbitrary $S$ and $N$. In Fig. 3 we
present the calculated $T$ as functions of $D/J$ for $N=1$, with $S$
taking nine values from 1 to 5. In Fig. 4 we present our accurate calculated
$T$ curves for $N$=1 through 4 and $S$=1, 3/2, 2, and 5/2.

\begin{figure}[!tbp]
  \includegraphics[width=7cm]{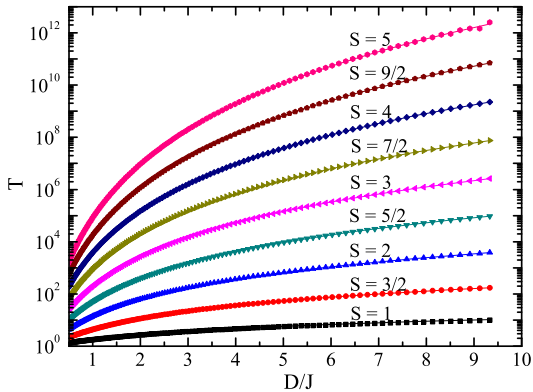}
\caption{The life times $T$ (in unit of $\hbar/J$) of the
two-spin chains as functions of $D/J$ for nine $S$ values, fitted
with Eq. (5).} \label{fig3}
\end{figure}

\begin{figure}[!htbp]
  \includegraphics[width=8.4cm]{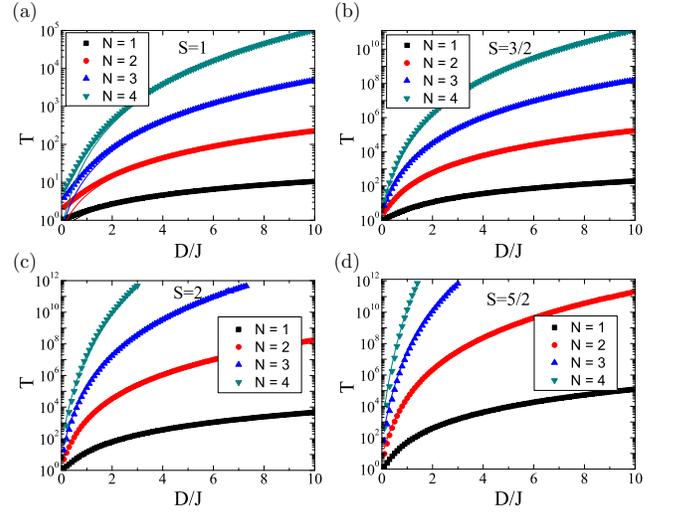}
\caption{The life times $T$ (in unit of $\hbar/J$) of the
$2N$ chains as function of $D/J$ for $S=1$
(a), $S=3/2$ (b), $S=2$ (c), and (d) $S=5/2$, fitted with Eq. (5).}
\label{fig4}
\end{figure}

It is very interesting that  all these ($D/J$)-$T$ curves can be
satisfactorily fitted with one simple function,
\begin{equation}
T = A \frac{\hbar}{J}(2\frac{D}{J}+1)^{N(2S-1)}
\end{equation}
where $A$ is a constant depending on $S$ and $N$ only. It is
surprising that, as we show in Fig. 5, $A$ can be well fitted with
$A=ba^N$, and furthermore the parameters $a$ and $b$ can be well fitted with
$a=0.2427\times 4.1545^S$ and $b=0.5007\times S^{-2.0713}$. The
fitted data of $A$, $a$, and $b$ are summarized in Table \ref{tab3}.
Consequently, we obtain a unified expression for $T$ as functions of
$D$, $J$, $N$, and $S$. It can be used to extrapolate $T$ with given
$D$ and $J$ for higher $S$ and larger $N$. It should be pointed out that
although $T$ increases with increasing $D$ or decreasing $J$, too
small $J$ will be harmful to stability against thermal fluctuations. Although $T$ increases exponentially with $N$ increasing, one cannot use too long spin chains for practical information storage because the N\'{e}el weight will decrease with $N$ increasing. Therefore, for a practically useful system, one should keep a balance between a
large $T$ and a good stability of the N\'{e}el states.

 \begin{table}[!h]\centering
 \caption{Fitted results of $A$, $a$, and $b$ for the antiferromagnetic spin-$S$ chains including  $2N$ spins.}\label{tab3}
\renewcommand{\multirowsetup}{\centering}
 \begin{tabular}{c|cccc|cc}
 \hline
   \hline
      \multirow{2}{*}{Spin}&\multicolumn{4}{c|}{$A$}&\multirow{2}{*}{$a$}&\multirow{2}{*}{$b$}\\
 \cline{2-5}
                &$N$=1  & $N$=2  & $N$=3 & $N$=4&&\\
   \hline
   $S$=1    &$0.506$  &$0.519$     &$0.531$     &$0.542$ & 1.023  & 0.4952    \\
   $S$=3/2  &$0.445$  &$0.909$     &$1.839$     &$3.725$ & 2.029  & 0.2202  \\
   $S$=2    &$0.500$  &$2.039$     &$8.527$     &$36.47$ & 4.178  & 0.1183  \\
   $S$=5/2  &$0.640$  &$5.245$     &$47.49$     &$371.9$ & 8.415  & 0.07596   \\
   $S$=3    &$0.894$  &$15.74$     &$289.2$     &4871    & 17.68  & 0.05080  \\
   \hline
   \hline
 \end{tabular}
 \end{table}

\begin{figure}[!tbp]
  \includegraphics[width=8.6cm]{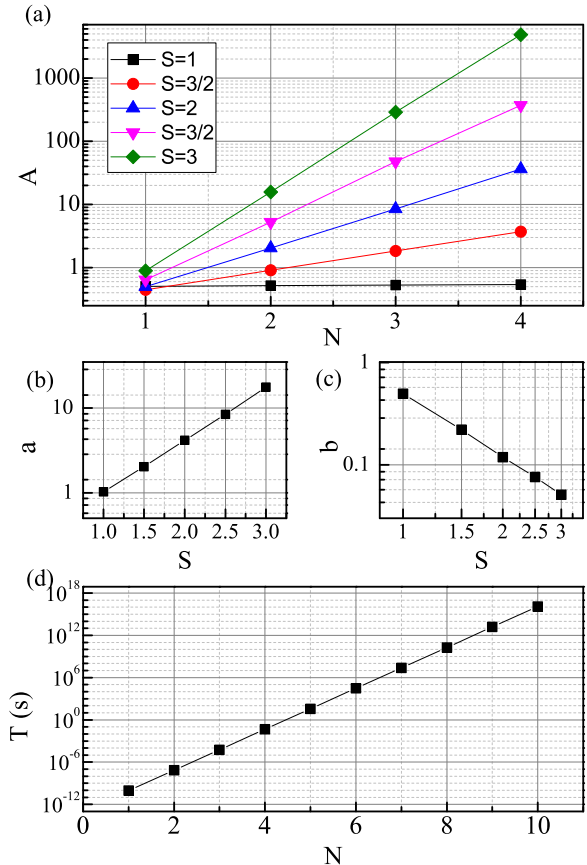}
\caption{(a) Fitting of the numeric results of $A$ depending on $N$ for different
$S$; (b) fitting of $a$ depending on $S$; (c) fitting of $b$ depending on $S$; and
(d) Plot of the life time $T$ (in second) depending on $N$  defined in Eq. (6) for
the Fe adatom spin chains on Cu$_2$N surface.} \label{fig5}
\end{figure}

\vspace{0.25cm}
{\noindent \large 2.4 Long life times in real adatom spin chains}
\vspace{0.25cm}

 For the short antiferromagnetic
chains of Fe adatom spins on Cu$_2$N surface, experimental result
reveals that $S=2$, $J=0.7$ meV, and $D=1.87$ meV \cite{AFM4}. In this case, we have $D/J=2.67$, belonging to the regime of strong uniaxial anisotropy. As a result, we
obtain a simple formula of $T$ (in second) depending on $N$,
\begin{equation}
T=1069^N\times 1.120\times 10^{-13}.
\end{equation}
We plot it in Fig. 5(d). The expression (6) implies that the switching rate ($1/T$) will be decreased approximately by 1000 times when we add two more Fe-adatom spins to the chain, which is consistent with the low-temperature limit of the experimental results\cite{ATM3}. This implies that the life times $T$ can be very long, reaching 1.9 days, 5.7
years, and 6057 years when $N$ is equivalent to 6, 7, and 8,
respectively. These results show that for such antiferromagnetic
chains, 14 or 16 spins (for $2N$) should be enough to achieve stable N\'{e}el
states at ultra-low temperatures (a few kelvin) for practical information storage.

On the other hand, for practical usage, we need to consider other factors affecting the
life times of N\'{e}el states. First, we consider possible transverse
single-ion anisotropy $E$ and transverse magnetic field $B_x$ which
appear  as additional
$\sum_{i}\{E[(\hat{S}_{i}^{x})^{2}-(\hat{S}_{i}^{y})^{2}]+\gamma
B_x\hat{S}_{i}^{x}\}$ in the Hamiltonian. Our calculations reveal
that as long as $D/J$ is not less than 1, there is little change in $T$ even
when $E/J$ and $\gamma B_x/J$ reach to 0.2. Then, we investigate
effect of spin exchange anisotropy on $T$, showing that the effect
is very small for $D/J> 1$. Therefore, our $T$ results are robust and
technically sound.

\vspace{0.5cm}
{\noindent \Large 3. Conclusions}
\vspace{0.5cm}

In summary, we have investigated the intrinsic quantum dynamics and life times of N\'{e}el states in the quantum Heisenberg antiferromagnetic chains with strong uniaixal single-ion anisotropy. For typical values of spin, chain length, and magnetic anisotropy, we have used exact diagonalization method to accurately calculate eigenvalues and eigenfunctions, and shown rigorously that the ground state and the first excitation are both safely separated from the other states and can be well described with the two N\'{e}el states as long as both $D/J$ and $S$ are large enough and $N$ is not too large. Through investigating accurate time evolution of the N\'{e}el states, we have determined their switching rates and life times. Surprisingly, we have achieved a unified powerful expression of the life times for arbitrary values of $N$, $S$, $D$, and $J$. Furthermore, we show that for the Fe-adatom spin antiferromagnets on Cu$_2$N semiconductor surface\cite{ATM1,ATM2,ATM3}, $2N$= 14 or 16 is large enough to achieve practically long life times for the N\'{e}el states. These theoretical results should be useful to help realize the N\'{e}el state storage of information and atomic-scale antiferromagnetic spintronics.

\vspace{0.5cm}
{\noindent \Large Acknowledgments}
\vspace{0.5cm}

This work is supported by Nature
Science Foundation of China (Grant Nos. 11174359 and 11574366), by Chinese
Department of Science and Technology (Grant No. 2012CB932302), and
by the Strategic Priority Research Program of the Chinese Academy of
Sciences (Grant No. XDB07000000).





\end{document}